# Performance Modelling and Analysis of Connection Admission Control in OFDMA based WiMAX System with MMPP Queuing

Abdelali EL BOUCHTI    Said EL KAFHALI    and    Abdelkrim HAQIQ
Computer, Networks, Mobility and modelling laboratory
e-NGN research group, Africa and Middle East
FST, Hassan 1st University, Settat, Morocco
a.elbouchti@gmail.com, kafhalisaid@gmail.com, ahaqiq@gmail.com

---

Abstract—This paper presents a problem of queuing theoretic performance modeling and analysis of Orthogonal Frequency Division Multiple Access (OFDMA) under broad-band wireless networks. We consider a single-cell WiMAX environment in which the base station allocates sub channels to the subscriber stations in its coverage area. The sub channels allocated to a subscriber station are shared by multiple connections at that subscriber station. To ensure the Quality of Service (QoS) performances, two Connection Admission Control (CAC) schemes, namely, threshold-based and queue-aware CAC schemes are considered at a subscriber station. A queuing analytical framework for these admission control schemes is presented considering OFDMA-based transmission at the physical layer. Then, based on the queuing model, both the connection-level and the packet-level performances are studied and compared with their analogues in the case without CAC. The connection arrival is modeled by a Poisson process and the packet arrival for a connection by a Markov Modulated Poisson Process (MMPP). We determine analytically and numerically different performance parameters, such as connection blocking probability, average number of ongoing connections, average queue length, packet dropping probability, queue throughput and average packet delay.

---

## I. Introduction

### *A. Reference system*

Wireless metropolitan area network (WMAN) technology based on IEEE 802.16 standard [3] and its evolutions has been developed to deliver a variety of multimedia services with different QoS requirements, such as throughput, delay, delay jitter, fairness and packet loss rate. Also known as Worldwide Interoperability for Microwave Access (WiMAX) [2]. As one of the most expected features, WiMAX [19] is supposed to have the capability of accommodating a variety of traffics, including data transfer, voice, and video. Correspondingly, four types of service, that is, Unsolicited Grant Service (UGS), real-time Polling Service (rtPS), non-real-time Polling Service (nrtPS), and Best Effort (BE) are defined by WiMAX Forum. Among them, UGS, rtPS, and nrtPS are classified into the category of QoS guaranteed services. To handle a multi service WiMAX access network of heterogeneous traffic load, the resource management scheme that can efficiently allocate radio and bandwidth resources to different subscribers and services is essential.

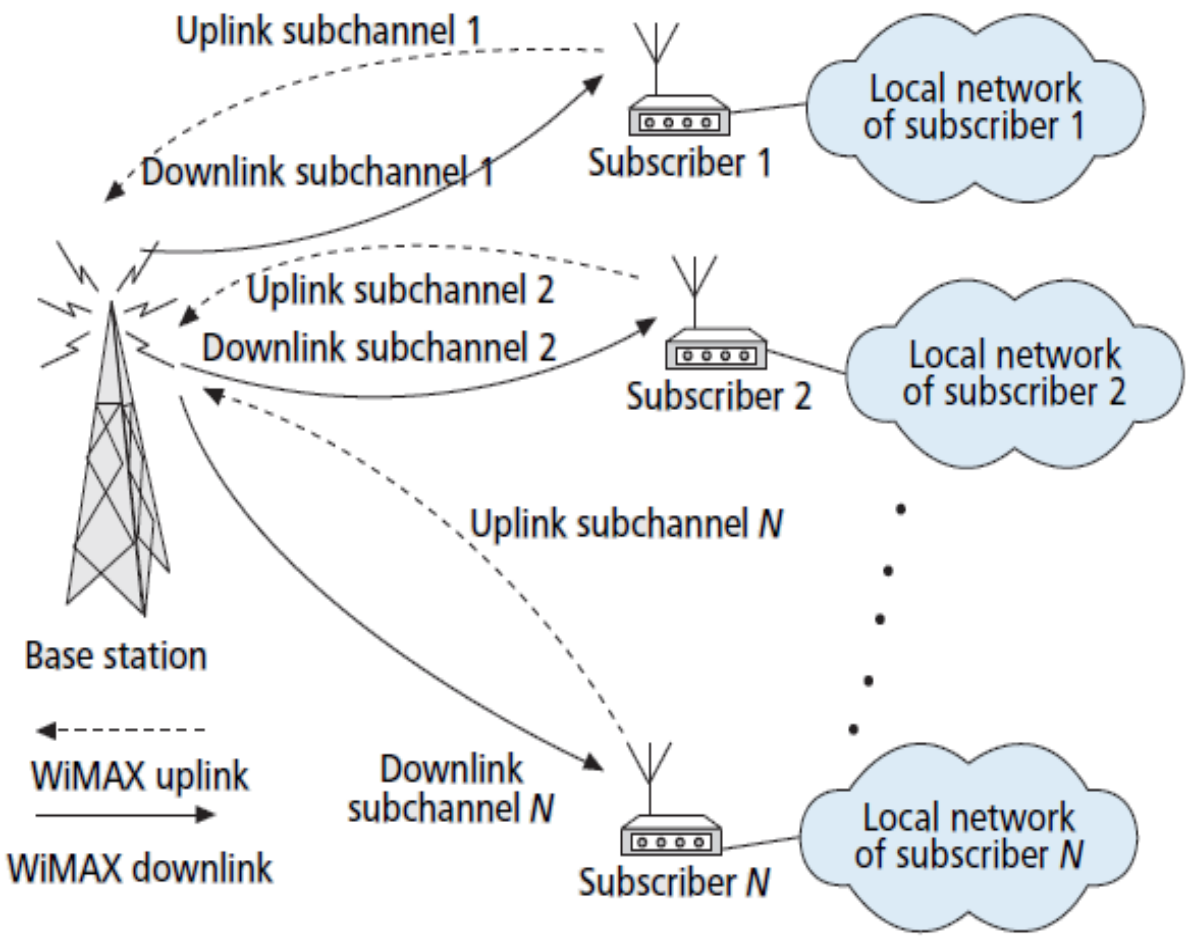

Figure 1. WiMAX point-to-multipoint network.

WiMAX technology supports both mesh and Point to Multi Point (PMP) networks. WiMAX mesh network is usually used for constructing wide-area wireless backhaul network, such as citywide wireless coverage and 3G RNC to base station connection. In the mean-time, WiMAX PMP network aims at providing the last-mile access to a broadband ISP. In this paper we only investigate the WiMAX PMP network. Figure 1 demonstrates a WiMAX PMP network of one base station and N subscribers. As standardized, WiMAX employs orthogonal frequency division multiplexing (OFDM) in its physical layer. Particularly, IEEE 802.16 has specified two flavours of the OFDM systems: one is simply identified as OFDM, and the other as OFDMA. With OFDMA, the wireless access performance can be substantially improved by transmitting data via multiple parallel channels, and also it is robust to inter-symbol interference and frequency-selective fading. OFDMA has been adopted as the physical layer transmission technology for IEEE 802.16/WiMAX-based broadband wireless networks. Although the IEEE 802.16/WiMAX standard [17] defines the physical layer specifications and the Medium Access Control (MAC) signalling mechanisms, the radio resource management methods such as those for CAC and dynamic bandwidth adaptation are left open. However, to guarantee QoS performances (e.g., call blocking rate, packet loss, and delay), efficient admission control is necessary in a WiMAX network at both the subscriber and the base stations.

### B. Related works and their drawbacks

In a wireless transmission system, arriving packets are buffered into radio link level queue before transmission to the target mobile (in downlink) or to the base station (in uplink). This buffering process causes delay in wireless transmission.
Again, the transmission rate of a wireless channel varies due to variations in channel quality and channel errors. Therefore, a queueing analysis [7] which can capture the channel and the radio link level buffer dynamics becomes a useful mathematical tool to investigate the QoS performances of a wireless transmission system. By using a queuing model, the performance results can be obtained more efficiently when compared with simulations. In addition, a queuing analytical model can be used not only to analyze the system's behaviour under different parameter settings, but also to optimize the system performances in which several standard techniques in optimization (e.g., Markov decision process [1],[5],[6]) can be applied.

The admission control problem was studied extensively for wired networks and also for traditional cellular wireless systems. The classical approach for CAC in a mobile wireless network is to use the guard channel scheme [10] in which a portion of wireless resources is reserved for handoff traffic. A more general CAC scheme, namely, the fractional guard scheme, was proposed [18] in which a handoff call/connection is accepted with a certain probability. To analyze various connection admission control algorithms, analytical models based on continuous-time Markov chain, were proposed [9]. However, most of these models dealt only with call/connection-level performances for the traditional voice-oriented cellular networks. In addition to the connection-level performances, packet-level (i.e., in-connection) performances also need to be considered for data-oriented packet-switched wireless networks such as WiMAX networks.

An earlier relevant work was reported by the authors in [15]. They considered a similar model in OFDMA based-IEEE 802.16 but they modelled both the connection-level and packet-level by tow different Poisson processes and they compared various QoS measures of CAC schemes. In [20], the authors proposed a Discrete-Time Markov Chain (DTMC) framework based on an MMPP traffic model to analyze VoIP performance. The MMPP processes are very suitable for formulating the multi-user VoIP traffic and capturing the inter-frame dependency between consecutive frames.

In addition, in [8] the authors studied the same model and they addressed the problem of queuing theoretic performance modelling and analysis of OFDMA transmission under admission control. They presented a connection admission control scheme for a multi-channel and multi-user OFDMA network and they modelled the connection arrival by a Poisson process and the packet arrival for a connection by a two-state MMPP process.

### C. Aims of the paper

In this paper, we present two CAC schemes for a multi-channel and multi-user OFDMA network. The first scheme is threshold-based, in which the concept of guard channel is used to limit the number of admitted connections to a certain threshold. The second scheme, namely, queue-aware is based on the information on queue status and it also inherits the concept of fractional guard channel in which an arriving connection is admitted with certain connection acceptance probability. Specifically, the connection acceptance probability is determined based on the queue status (i.e., the number of packets in the queue). A queuing analytical model is developed based on a three-DTMC which captures the system dynamics in terms of the number of connections and queue status. We assume that the connection arrival and the packet arrival for a connection follow a Poisson process and an MMPP process respectively. Based on this model, various performance parameters such as the connection blocking probability, the average number of ongoing connections, the average queue length, the probability of packet dropping due to lack of buffer space, the queue throughput, and the average queuing delay are obtained. The numerical results reveal the comparative performance characteristics of the threshold-based and the queue-aware CAC algorithms in an OFDMA-based WiMAX network

### D. Organisation of the paper

This paper is structured as follows. Section II describes the system model including the objective of CAC policy. The formulation of the analytical model for connection admission control is presented in section III. In section IV we determine analytically different performance parameters. Numerical results are stated in section V. Finally, section VI concludes the paper.

## II. MODEL DESCRIPTION

### A. System model

We consider a single cell in a WiMAX network with a base station and multiple subscriber stations (Figure 2). Each subscriber station serves multiple connections. Admission control is used at each subscriber station to limit the number of ongoing connections through that subscriber station. At each subscriber station, traffic from all uplink connections are aggregated into a single queue [16]. The size of this queue is finite (i.e., $X$ packets) in which some packets will be dropped if the queue is full upon their arrivals. The OFDMA transmitter at the subscriber station retrieves the head of line packet(s) and transmits them to the base station. The base station may allocate different number of sub channels to different subscriber stations. For example, a subscriber station with higher priority could be allocated more number of sub channels.

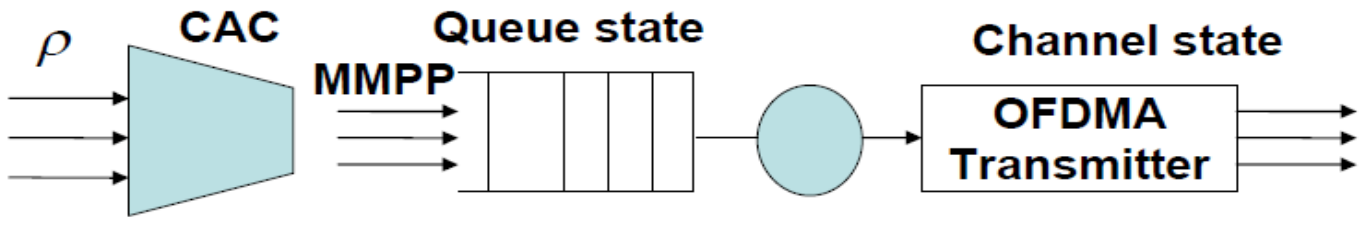

Figure 2. System model

*B. CAC Policy*

The main objective of a CAC mechanism is to limit the number of ongoing connections/flows so that the QoS performances can be guaranteed for all the ongoing connections. Then, the admission control decision is made to accept or reject an incoming connection. To ensure the QoS performances of the ongoing connections, the following CAC schemes for subscriber stations are proposed.

*B.1 Threshold-Based CAC*

In this case, a threshold C is used to limit the number of ongoing connections. When a new connection arrives, the CAC module checks whether the total number of connections including the incoming one is less than or equal to the threshold C. If it is true, then the new connection is accepted, otherwise it is rejected.

*B.2 Queue-Aware CAC*

This scheme works based on connection acceptance probability $\alpha_x$ which is determined based on the queue status. Specifically, when a connection arrives, the CAC module accepts the connection with probability $\alpha_x$ , where $x\ (x \in \{0,1,...,X\})$ is the number of packets in the queue in the current time slot. Here, $X$ denotes the size of the queue of the subscriber station under consideration. Note that the value of the parameter $\alpha_x$ can be chosen based on the radio link level performance (e.g., packet delay, packet dropping probability) requirements.

## III. Formulation of the Analytical Model

*A. Formulation of the Queueing Model*

An analytical model based on DTMC is presented to analyze the system performances at both the connection-level and at the packet-level for the connection admission control schemes described before. To simplify the analytical model, we assume that packet arrival for a connection follows an MMPP process [4] which is identical for all connections in the same queue. The connection inter-arrival time and the duration of a connection are assumed to be exponentially distributed with average $1/\rho$ and $1/\mu$, respectively.

An MMPP is a stochastic process in which the intensity of a Poisson process is defined by the states of a Markov chain. That is, the Poisson process can be modulated by a Markov chain. As mentioned before, an MMPP process can be used to model time-varying arrival rates and can capture the inter-frame dependency between consecutive frames ([11], [12], [13]). The transition rate matrix and the Poisson arrival rate matrix of the MMPP process can be expressed as follows:

$$Q_{MMPP} = \begin{pmatrix} \delta_{11} & \cdots & \delta_{1S} \\ \vdots & \ddots & \vdots \\ \delta_{S1} & \cdots & \delta_{SS} \end{pmatrix}, \quad \Lambda = \begin{pmatrix} \lambda_1 & & \\ & \ddots & \\ & & \lambda_S \end{pmatrix} \tag{1}$$

$\lambda_s$ denotes the arrival rate at state (phase) $s$ and $\delta_{ss'}$ is the transition rate from $s$ to $s'$ *(s, s'=1, 2, ..., S)*. The steady-state probabilities of the underlying Markov chain are given by:

$$\pi_{MMPP} = \pi_{MMPP}.Q_{MMPP} \text{ and } \pi_{MMPP}.1 = 1 \tag{2}$$

where $1$ is a column matrix of ones.

The mean steady state arrival rate generated by the MMPP is:

$$\lambda_{MMPP} = \pi_{MMPP}\lambda^T \tag{3}$$

where $\lambda^T$ is the transpose of the row vector $\lambda = (\lambda_1, ......, \lambda_S)$. The state of the system is described by the process $Y_t = (S_t, Y_t^1, Y_t^2)$, where $S_t$ is the state (phase) of an irreducible continuous time Markov chain and $Y_t^1$ (respectively $Y_t^2$) is the number of packets in the aggregated queue (the number of ongoing connections) at the end of every time slot $t$.

Thus, the state space of the system for both the CAC schemes is given by:

$$E = \{(s, x, c) / s \in \{1, ...., S\},\ 0 \le x \le X,\ c \ge 0\}.$$

For the both CAC algorithms, the number of packet arrivals depends on the number of connections. However, for the queue-aware CAC algorithm, the number of packets in the queue affects the acceptance probability for a new connection. The state transition diagram is shown in figure 3.

Note that the probability that $n$ Poison events with average rate $\rho$ occur during an interval T can be obtained as follows:

$$f_n(\rho) = \frac{e^{-\rho T}(\rho T)^n}{n!} \tag{4}$$

This function is required to determine the probability of both connection and packet arrivals.

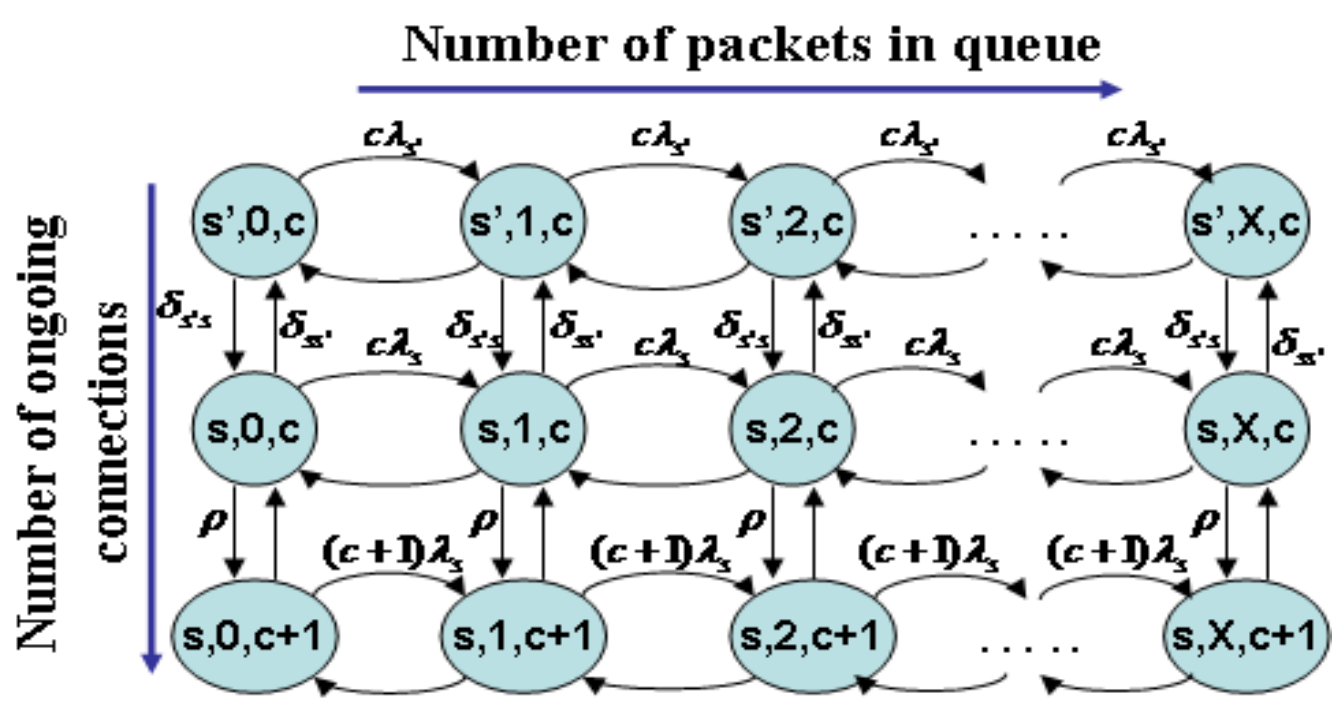

Figure 3. State transition diagram of discrete time Markov chain.

### B. Threshold-Based CAC Algorithm

In this case, the transition matrix $Q$ for the number of connections in the system can be expressed as follows:

$$Q = \begin{bmatrix} q_{0,0} & q_{0,1} & & & \\ q_{1,0} & q_{1,1} & q_{1,2} & & \\ \ddots & \ddots & \ddots & & \\ & & q_{C-1,C-2} & q_{C-1,C-1} & q_{C-1,C} \\ & & & q_{C,C-1} & q_{C,C} \end{bmatrix} \quad (5)$$

where each row indicates the number of ongoing connections. As the length of a frame $T$ is very small compared with connection arrival and departure rates, we assume that the maximum number of arriving and departing connections in a frame is one. Therefore, the elements of this matrix can be obtained as follows:

$$\begin{aligned} q_{c,c+1} &= f_1(\rho)\times(1-f_1(c\mu)), \quad \text{c=0,1,...,C-1} \\ q_{c,c-1} &= (1-f_1(\rho))\times f_1(c\mu), \quad \text{c=1,2,...,C} \\ q_{c,c} &= f_1(\rho)\times f_1(c\mu)+(1-f_1(\rho))\times(1-f_1(c\mu)), \quad \text{c=0,1,...,C} \end{aligned} \quad (6)$$

where $q_{c,c+1}$, $q_{c,c-1}$ and $q_{c,c}$ represent the cases that the number of ongoing connections increases by one, decreases by one, and does not change, respectively.

### C. Queue-Aware CAC Algorithm

Because the admission of a connection in this case depends on the current number of packets in the queue, the transition matrix can be expressed based on the number of packets ($x$) in the queue as follows:

$$Q_x = \begin{bmatrix} q_{0,0}^{(x)} & q_{0,1}^{(x)} & \\ q_{1,0}^{(x)} & q_{1,1}^{(x)} & q_{1,2}^{(x)} \\ q_{2,0}^{(x)} & q_{2,1}^{(x)} & q_{2,2}^{(x)} \\ \ddots & \ddots & \ddots \end{bmatrix} \quad (7)$$

where:

$$\begin{aligned} q_{c,c+1}^{(x)} &= f_1(\alpha_x\rho)\times(1-f_1(c\mu)), \quad \text{c=0,1,...} \\ q_{c,c-1}^{(x)} &= (1-f_1(\alpha_x\rho))\times f_1(c\mu), \quad \text{c=1,2,...} \\ q_{c,c}^{(x)} &= f_1(\alpha_x\rho)\times f_1(c\mu)+(1-f_1(\alpha_x\rho))\times(1-f_1(c\mu)), \quad \text{c=0,1,...} \end{aligned} \quad (8)$$

in which $\alpha_x$ is the connection acceptance probability when there are x packets in the queue.

### D. Transition Matrix for the Queue

The transition matrix $P$ of the entire system can be expressed as in equation 7. The rows of matrix $P$ represent the number of packets ($x$) in the queue.

$$P = \begin{bmatrix} p_{0,0} & \cdots & p_{0,A} & & & \\ \vdots & \vdots & \ddots & \ddots & & \\ p_{R,0} & \cdots & p_{R,R} & \cdots & p_{R,R+A} & \\ \ddots & \ddots & \ddots & \ddots & \ddots & \ddots \\ & p_{x,x-R} & \cdots & p_{x,x} & \cdots & p_{x,x+R} \\ & \ddots & \ddots & \ddots & \ddots & \ddots \end{bmatrix} \quad (9)$$

Matrices $p_{x,x'}$ represent the changes in the number of packets in the queue (i.e., the number of packets in the queue changing from $x$ in the current frame to $x'$ in the next frame). We first establish matrices $\mathrm{v}_{(s,x),(s,x')}$, where the diagonal elements of these matrices are given in equation 8. For $r\in\{0,1,2,...,D\}$ and $n\in\{0,1,2,...,(c\times A)\}$, $l=1,2,...,D$, and $m=1,2,...,(c\times A)$. The non-diagonal elements of $\mathrm{v}_{(s,x),(s,x')}$ are all zero.

$$\begin{aligned} \left[\mathrm{v}_{(s,x);(s,x-l)}\right]_{c,c} &= \sum_{n-r=l} f_n((c-1)\lambda_s)[\mathrm{R}]_r \\ \left[\mathrm{v}_{(s,x);(s,x+m)}\right]_{c,c} &= \sum_{r-n=m} f_n((c-1)\lambda_s)[\mathrm{R}]_r \\ \left[\mathrm{v}_{(s,x);(s,x)}\right]_{c,c} &= \sum_{r=n} f_n((c-1)\lambda_s)[\mathrm{R}]_r \end{aligned} \quad (10)$$

Note that, matrix **R** [15] has size 1 × $R$ + 1, where $R$ indicates the maximum number of packets that can be transmitted in one frame. Here, $A$ is the maximum number of packets that can arrive from one connection in one frame and $D$ is the maximum number of packets that can be transmitted in one frame by all of the allocated sub channels allocated to that particular queue and it can be obtained from $D=\min\,(R,x)$. This is due to the fact that the maximum number of transmitted packets depends on the number of packets in the queue and the maximum possible number of transmissions in one frame. Note that, $\left[\mathrm{v}_{(s,x);(s,x-l)}\right]_{c,c}$, $\left[\mathrm{v}_{(s,x);(s,x+m)}\right]_{c,c}$ and $\left[\mathrm{v}_{(s,x);(s,x)}\right]_{c,c}$ represent the probability that the number of packets in the queue increases by $l$, decreases by $m$, and does not change, respectively, when there are *c-1* ongoing connections. Here, $[\mathrm{v}]_{i,j}$ denotes the element at row $i$ and column $j$ of matrix v, and these elements are obtained based on the assumption that the packet arrivals for the ongoing connections are independent of each other.

Finally, we obtain the matrices $p_{x,x'}$ by combining both the connection-level and the queue-level transitions as follows:

$$p_{x,x'} = Q\mathrm{v}_{(s,x),(s,x')} \tag{11}$$

$$p_{x,x'} = Q_x\mathrm{v}_{(s,x),(s,x')} \tag{12}$$

for the cases of threshold-based (Equation 9) and queue-aware (Equation 10) CAC algorithms, respectively.

## IV. Performance Parameters

In this section, we determine the connection-level and the packet-level performance parameters (i.e., connection blocking probability, average number of ongoing connections in the system, and average queue length) for the both CAC schemes.

For the threshold-based CAC scheme, all of the above performance parameters can be derived from the steady state probability vector of the system states $\pi$, which is obtained by solving $\pi P = \pi$ and $\pi 1 = 1$, where $1$ is a column matrix of ones.

However, for the queue-aware CAC algorithm, the size of the matrix $Q_x$ needs to be truncated at $C_{tr}$ (i.e., the maximum number of ongoing connections at the subscriber station).

Also, the size of the matrix $P$ needs to be truncated at $X$ (i.e., the maximum number of packets in the queue) for the both schemes.

The steady-state probability, denoted by $\pi(s,x,c)$ for the state that there are $c$ connections and $x \in \{0,1,...,X\}$ packets in the queue, can be extracted from matrix $\pi$ as follows $\pi(s,x,c) = [\pi]_{s\times x\times((C+1)+c)}$, $s = 1,...,S;\ c = 0,1,...,C'$ (13) where $C' = C$ and $C' = C_{tr}$ for the threshold-based and the queue-aware CAC algorithms, respectively. Using these steady state probabilities, the various performance parameters can be obtained. Note that, the subscripts *tb* and *qa* are used to indicate the performance parameters for the threshold-based and the queue-aware CAC schemes, respectively.

### A. Connection Blocking Probability

This performance parameter indicates that an arriving connection will be blocked due to the admission control decision. It indicates the accessibility of the wireless service, and for the threshold-based CAC scheme. It can be obtained as follows:

$$p^{tb}_{block} = \sum_{s=1}^{S}\sum_{x=0}^{X}\pi(s,x,C). \tag{14}$$

The above probability refers to the probability that the system serves the maximum allowable number of ongoing connections. The blocking probability for the queue-aware CAC is obtained from

$$p^{qa}_{block} = \sum_{s=1}^{S}\sum_{x=0}^{X}\sum_{c=1}^{C_{tr}}((1-\alpha_x).\pi(s,x,C)). \tag{15}$$

in which the blocking probability is the sum of the probabilities of rejection for all possible number of packets in the queue.

### B. Average Number of Ongoing Connections

It can be obtained as

$$N^{tb}_{c} = \sum_{s=1}^{S}\sum_{x=0}^{X}\sum_{c=0}^{C} c.\pi(s,x,c) \tag{16}$$

$$N^{qa}_{c} = \sum_{s=1}^{S}\sum_{x=0}^{X}\sum_{c=0}^{C_{tr}} c.\pi(s,x,c) \tag{17}$$

### C. Average Queue Length Average

It is given by

$$N^{tb}_{x} = \sum_{s=1}^{S}\sum_{x=0}^{C}\sum_{x=0}^{X} x.\pi(s,x,c) \tag{18}$$

$$N^{qa}_{x} = \sum_{s=1}^{S}\sum_{x=0}^{C_{tr}}\sum_{x=0}^{X} x.\pi(s,x,c) \tag{19}$$

### D. Packet Dropping Probability

This performance parameter refers to the probability that an incoming packet will be dropped due to the unavailability of buffer space. It can be derived from the average number of dropped packets per frame. Given that there are $x$ packets in the queue and the number of packets in the queue increases by $m$, the number of dropped packets is $m-(X-x)$ for $m > X-x$, and zero otherwise. The average number of dropped packets per frame is obtained as follows:

$$N_{drop} = \sum_{s=1}^{S}\sum_{c=1}^{C}\sum_{x=0}^{X}\sum_{m=X-x+1}^{A}\left(\sum_{l=1}^{C}[p_{x,x+m}]_{c,l}\right).(m-(X-x)).\pi(s,x,c) \tag{20}$$

where the term $\left(\sum_{l=1}^{C}[p_{x,x+m}]_{c,l}\right)$ indicates the total probability that the number of packets in the queue increases by $m$ at every arrival phase. Note that, we consider probability $p_{x,x+m}$ rather than the probability of packet arrival as we have to consider the packet transmission in the same frame as well.

After calculating the average number of dropped packets per frame, we can obtain the probability that an incoming packet is dropped as follows:

$$p_{drop} = \frac{N_{drop}}{\bar{\lambda}} \tag{21}$$

where $\bar{\lambda}$ is the average number of packet arrivals per frame and it can be obtained from

$$\bar{\lambda} = \lambda_{MMPP} N_c. \tag{22}$$

### E. Queue throughput

It measures the number of packets transmitted in one frame and can be obtained from

$$\varphi = \lambda_{MMPP}(1 - p_{drop}). \tag{23}$$

### F. Average Packet Delay

It is defined as the number of frames that a packet waits in the queue since its arrival before it is transmitted. We use Little's law [14] to obtain average delay as follows:

$$D = \frac{N_x}{\varphi} \tag{24}$$

## V. NUMERICAL RESULTS

In this section we present the numerical results of both CAC schemes. We use the Matlab software to solve numerically and to evaluate the various performance parameters.

### A. Performance Evaluation

We consider one queue (which corresponds to a particular subscriber station) for which five sub-channels are allocated and we assume that the average SNR is the same for all of these sub-channels. Each sub-channel has a bandwidth of 160 kHz. The length of a sub frame for downlink transmission is one millisecond, and therefore, the transmission rate in one sub channel with rate ID = 0 (i.e., BPSK modulation and coding rate is 1/2) is 80 kbps. We assume that the maximum number of packets arriving in one frame for a connection is limited to 50.

For the threshold-based CAC mechanism, the value of the threshold C is varied according to the evaluation scenarios. For the queue-aware CAC mechanism, the value of the connection acceptance probability is determined as follows:

$$\alpha_{x=} \begin{cases} 1, & 0 \le x \prec B_{th} \\ 0, & B_{th} \le x \le X. \end{cases} \tag{25}$$

In the performance evaluation, we use $B_{th} = 90$.

For performance comparison, we also evaluate the queuing performance in the absence of CAC mechanism. For the case without CAC, we truncate the maximum number of ongoing connections at 40. The average duration of a connection is set to twenty minutes for all the evaluation scenarios. The queue size is 250 packets. The parameters are set as follows: The connection arrival rate is 0.8 connections per minute. When the underlying Markov chain is at state *s* (*s*=1, 2, 3, 4), we consider the packet arrival rate per connection is (5, 3, 15, 1) packet(s) per frame for each state *s* respectively. Average SNR on each sub-channel is 5 dB. For clarity, the all numerical parameters are summarized in table 1.

| | |
|---|---|
| $X$ | 250 |
| $A$ | 50 |
| $C_{tr}$ | 40 |
| $B_{th}$ | 90 |
| $\rho$ | 0,8 |
| $(\lambda_1, \lambda_2, \lambda_3, \lambda_4)$ | (5, 3, 15, 1) |
| $\mu$ | 20 |
| Average SNR | 5 dB |

TABLE I. SUMMARY OF NUMERICAL PARAMETERS.

Note that, we vary some of these parameters depending on the evaluation scenarios whereas the others remain fixed.

### B. Performance Analysis

We first examine the impact of connection arrival rate on connection-level performances. Variations in average number of ongoing connections and connection blocking probability with connection arrival rate are shown in Figures 4 and 5, respectively. As expected, when the connection arrival rate increases, the number of ongoing connections and connection blocking probability increase.

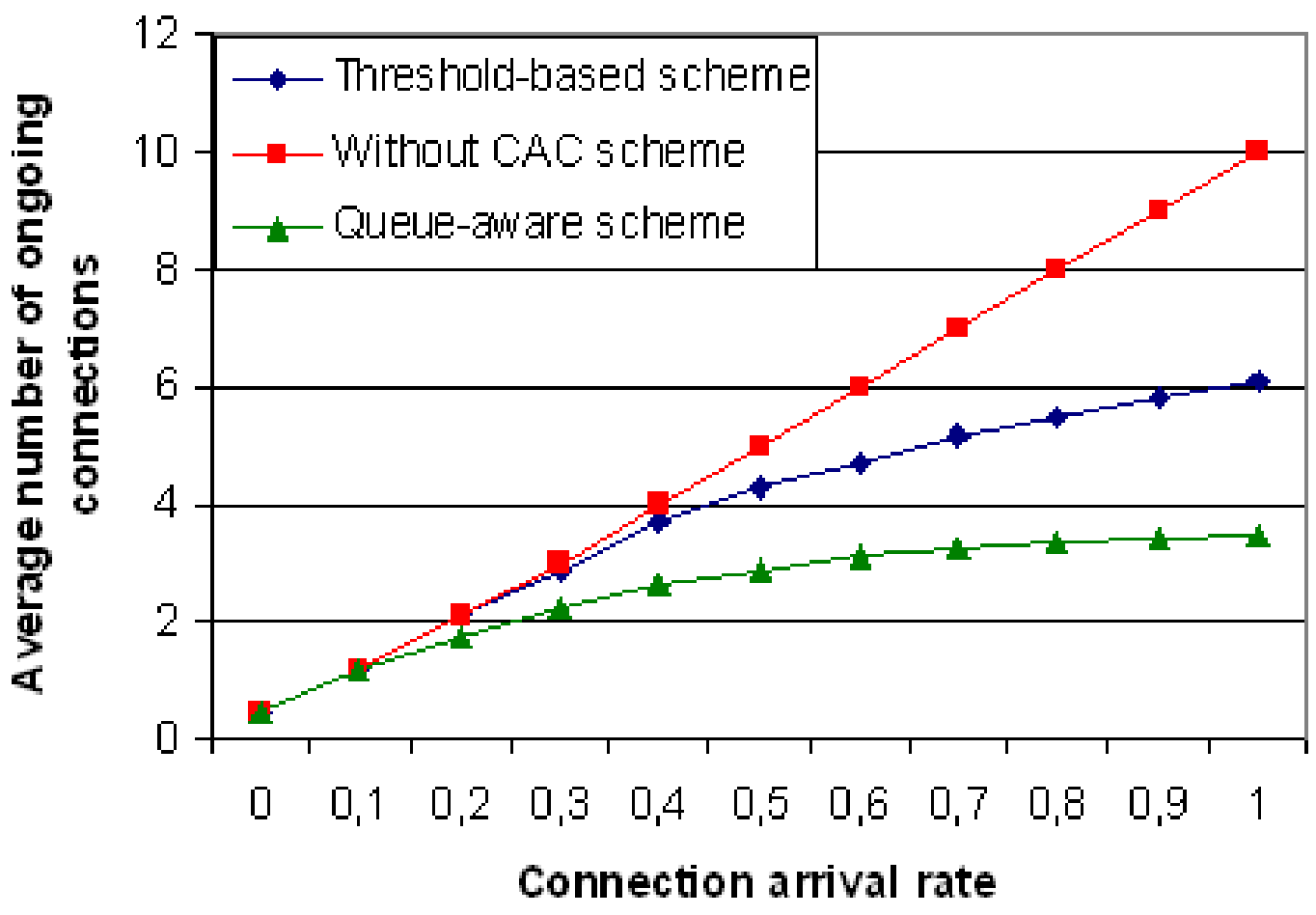

Figure 4: Average number of ongoing connections under different connection arrival rates.

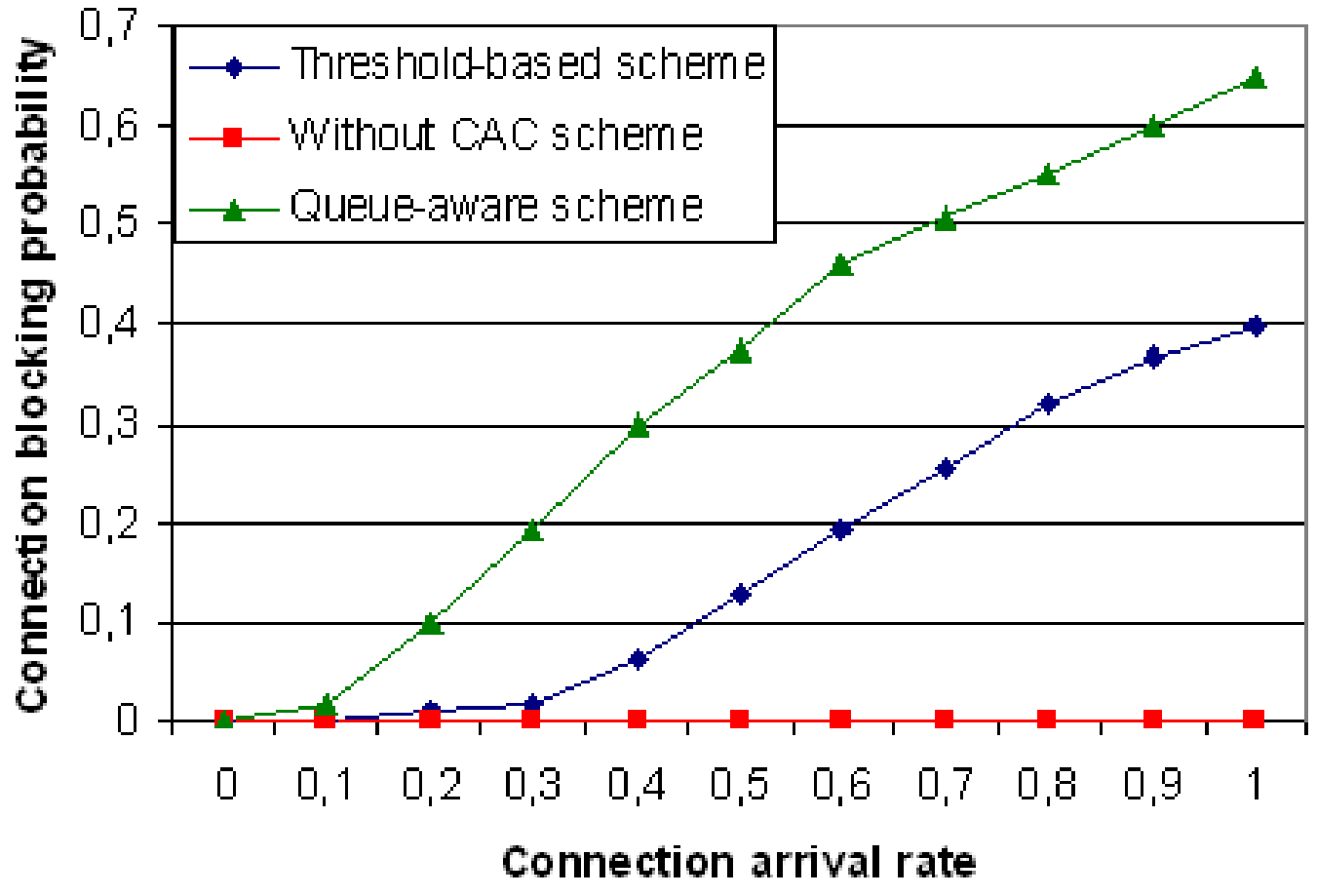

Figure 5: Connection blocking under different connection arrival rates.

The packet-level performances under different connection arrival rates are shown in Figures 6 through 9 for average number of packets in the queue, packet dropping probability, queue throughput, and average queueing delay, respectively. These performance parameters are significantly impacted by the connection arrival rate. Because the both CAC schemes limit the number of ongoing connections, packet-level performances can be maintained at the target level. In this case, both CAC schemes result in better packet-level performances compared with those without CAC scheme.

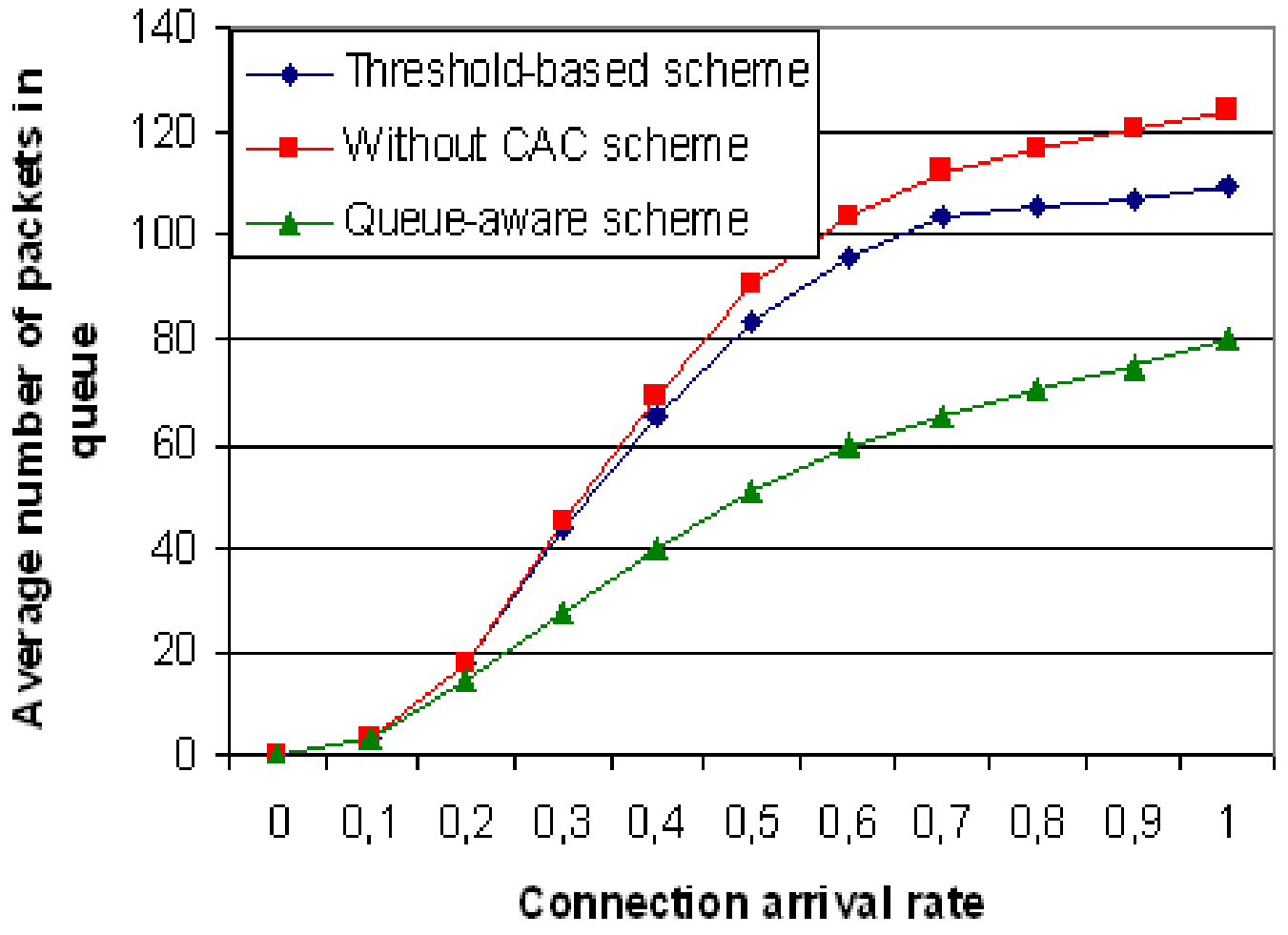

Figure 6: Average number of packets in queue under different connection rates.

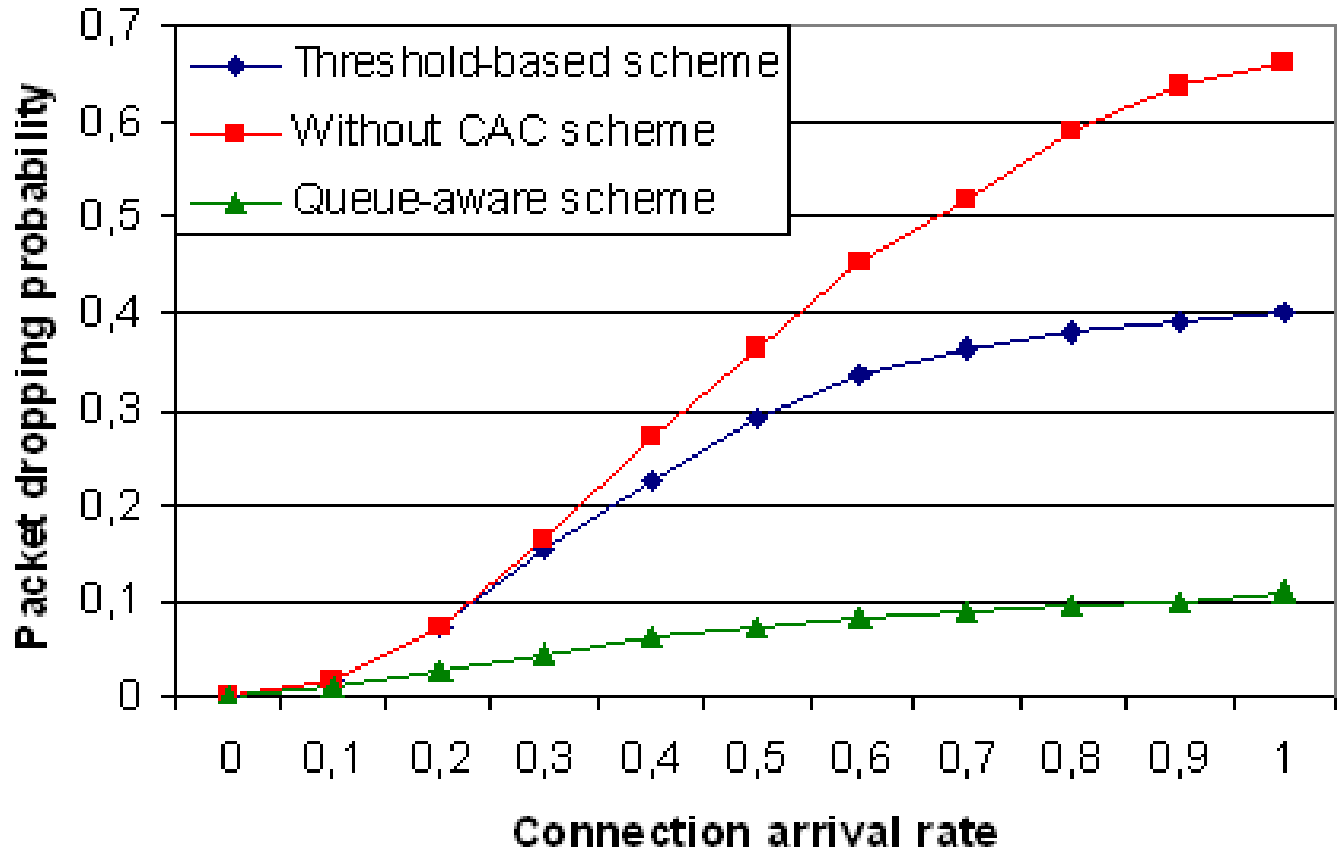

Figure 7: Packet dropping under different connection arrival rates.

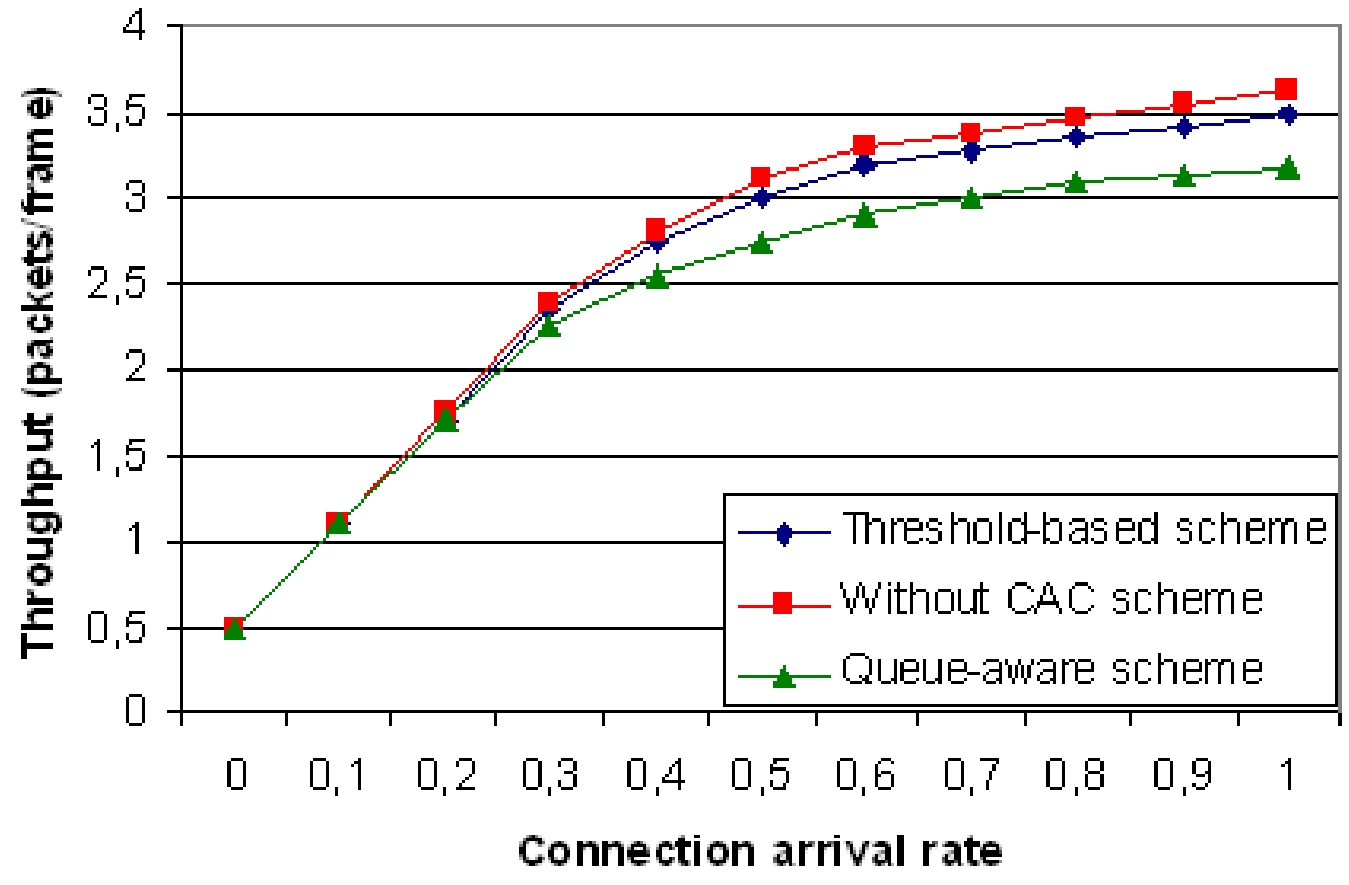

Figure 8: Queuing throughput under different connection arrival rates.

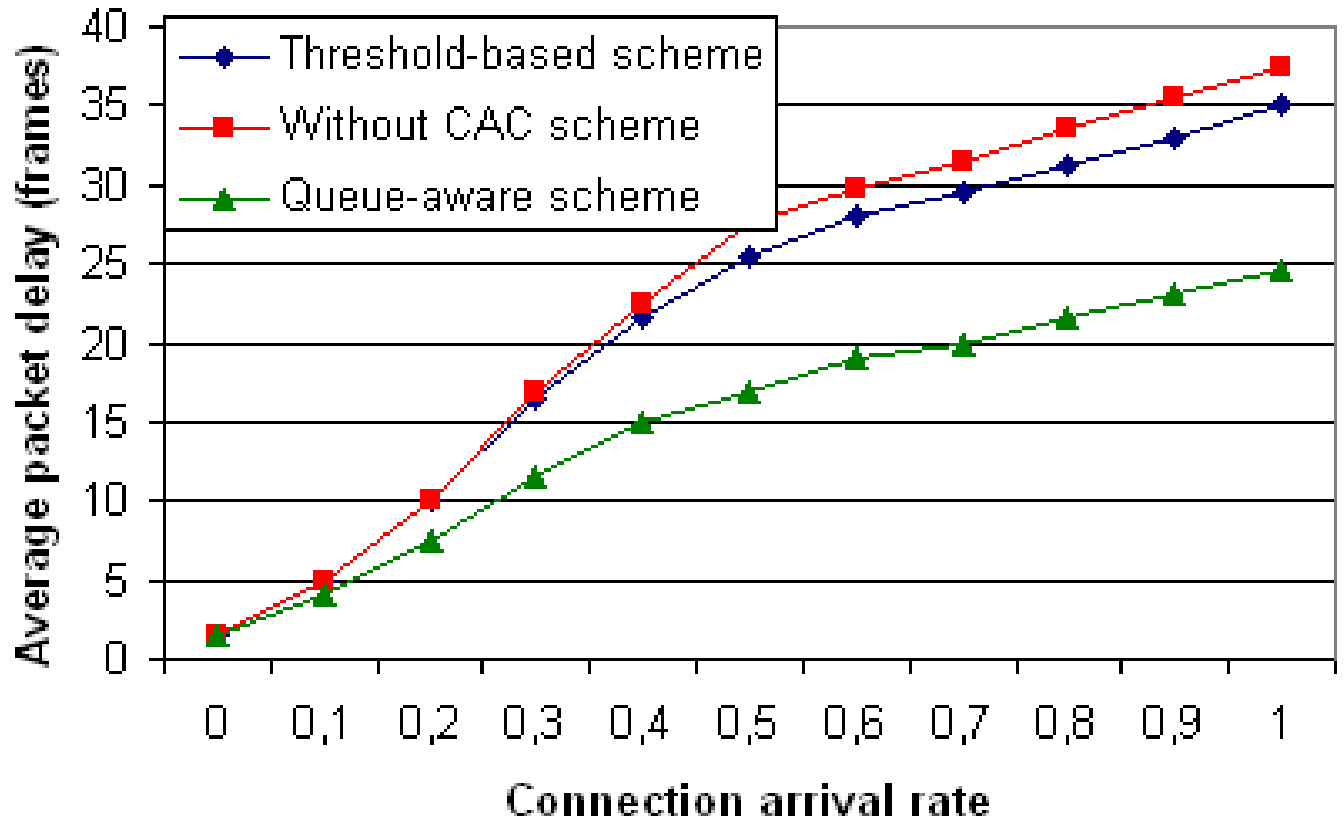

Figure 9: Average packet delay under different connection arrival rates.

Variations in packet dropping probability and average packet delay with channel quality are shown in Figures 10 and 11, respectively. As expected, the packet-level performances become better when channel quality becomes better. Also, we observe that the connection-level performances for the threshold-based CAC scheme and those without CAC scheme are not impacted by the channel quality when this later becomes better (the connection blocking probability remains constant when the channel quality varies), connection blocking probability decreases significantly for the queue-aware CAC mechanism when the channel quality becomes better (Figure. 12).

Based on these observations, we can conclude that the queue-aware CAC can adapt the admission control decision based on the queue status which is desirable for a system with high traffic fluctuations.

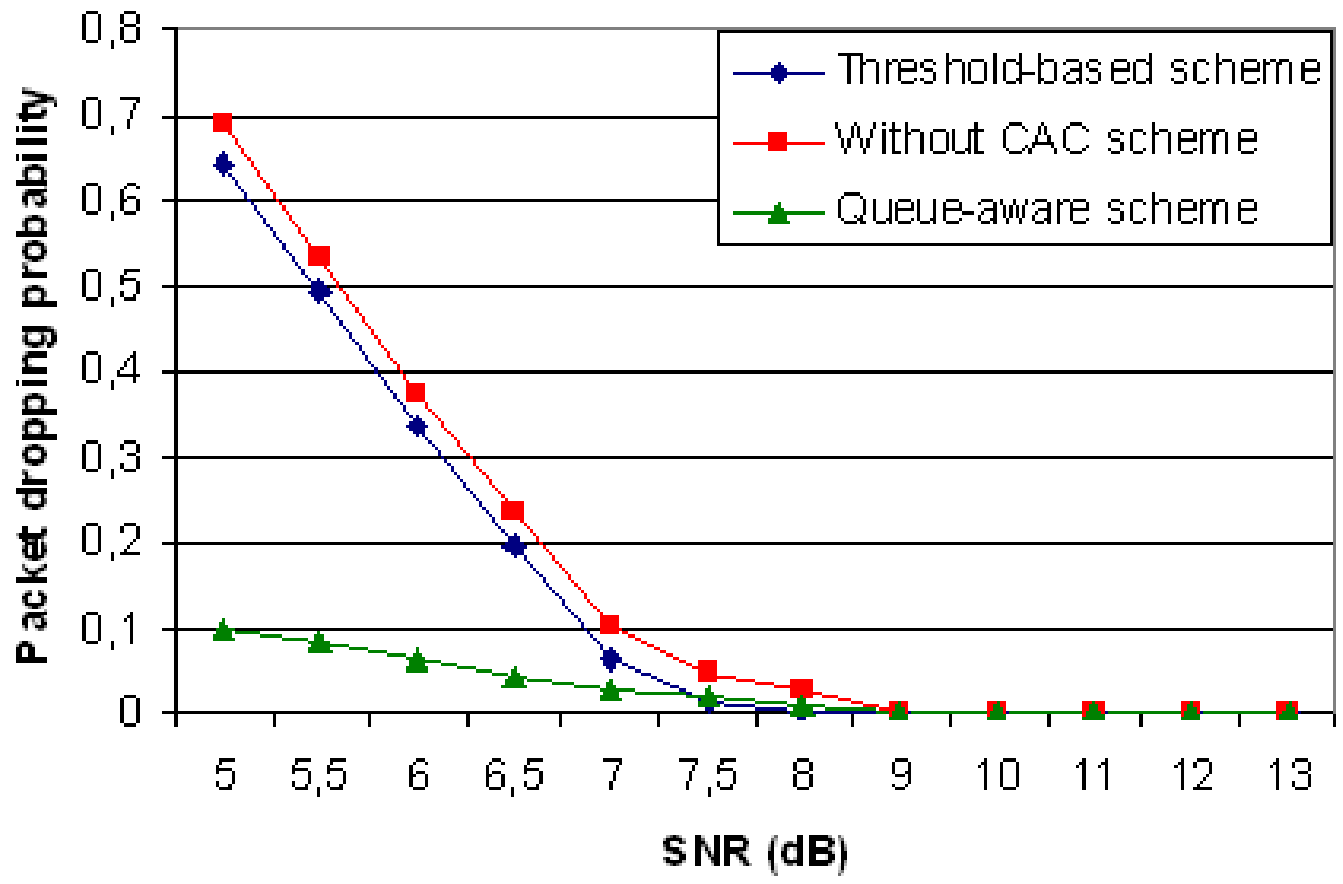

Figure 10: Packet dropping probability under different channel qualities.

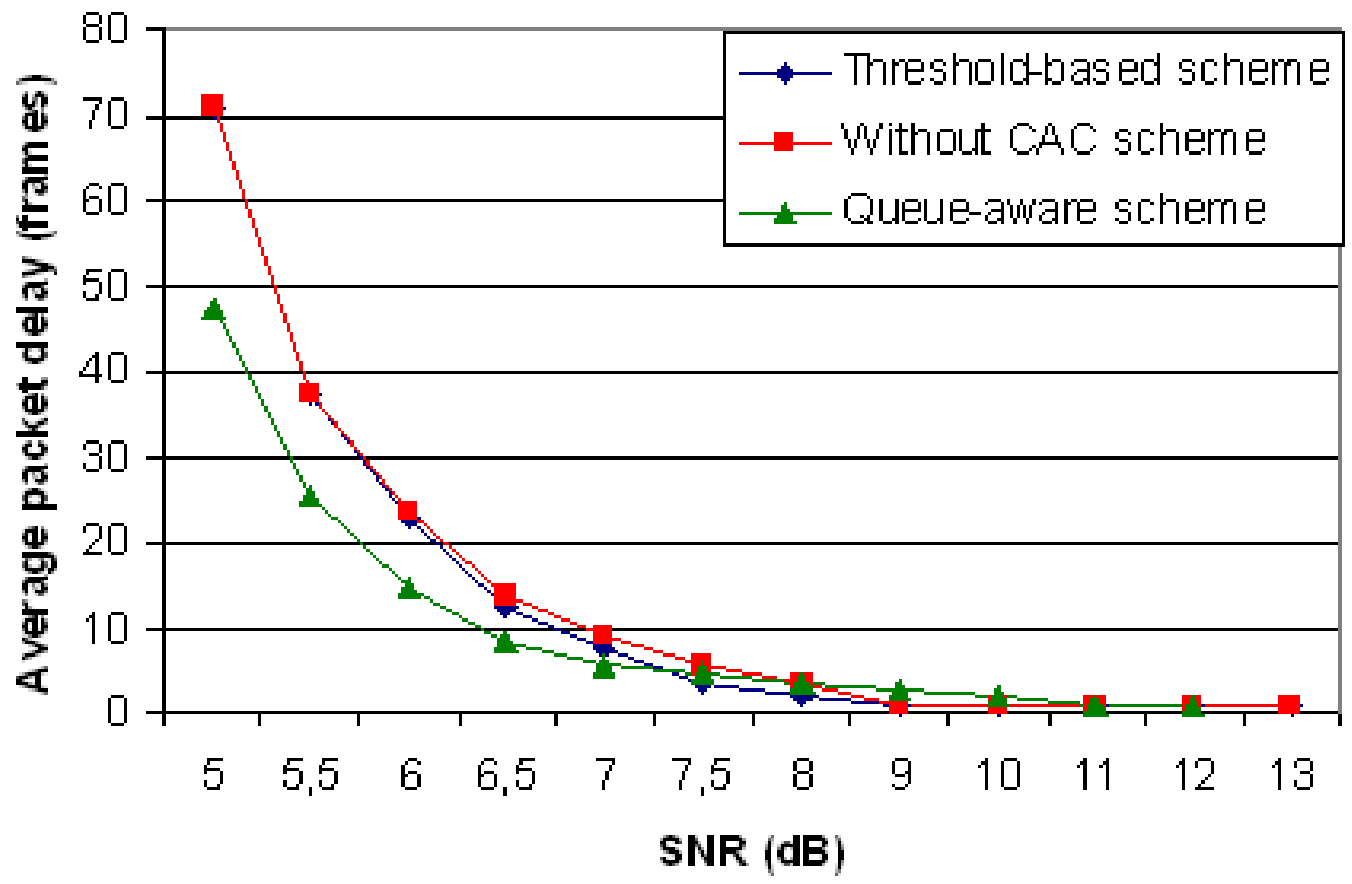

Figure 11: Average packet delay under different channel qualities.

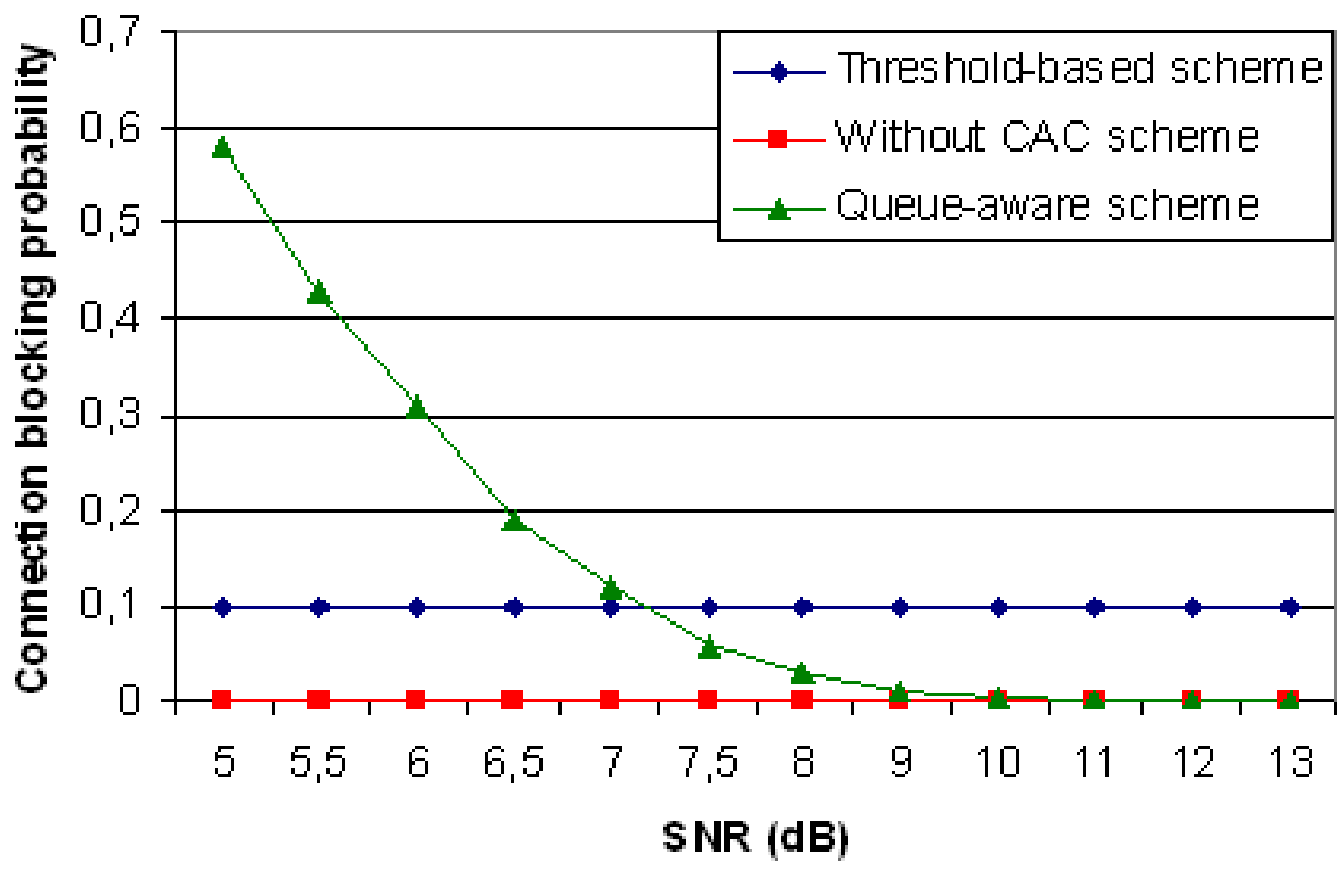

Figure 12: Connection blocking probability under different channel qualities.

## VI. Conclusion

In this paper, we have addressed the problem of queuing theoretic performance modelling and analysis of OFDMA transmission under admission control. We have considered a WiMAX system model in which a base station serves multiple subscriber stations and each of the subscriber stations is allocated with a certain number of sub channels by the base station. There are multiple ongoing connections at each subscriber station.

For CAC, we have considered two schemes, namely, queue-aware scheme and threshold-based scheme. While the threshold-based CAC scheme simply fixes the number of ongoing connections, the queue-aware CAC scheme considers the number of packets in the queue for the admission control decision of a new connection. The connection-level and packet-level performances of these CAC schemes have been studied based on the queuing model.

The connection-level and packet-level performances of the both CAC schemes have been studied based on the queuing model. The connection arrival is modelled by a Poisson process and the packet arrival for a connection by an MMPP process with $s$ ($s > 1$) phases. We have determined analytically and numerically different performance parameters, such as the connection blocking probability, the average number of ongoing connections, the average queue length, the packet dropping probability, the queue throughput and the average packet delay.

Numerical results in the case where $s$=4 show that, the performance parameters of connection-level and packet-level are significantly impacted by the connection-level rate, the both CAC schemes result in better packet-level performances compared with those without CAC scheme. The packet-level performances become better when channel quality becomes better. On the other hand, the connection-level performances for the threshold-based CAC scheme and those without CAC scheme are not impacted by the channel quality when this later becomes better. Then, the queue-aware CAC can adapt the admission control decision based on the queue status which is desirable for a system with high traffic fluctuations.

## AUTHORS PROFILE

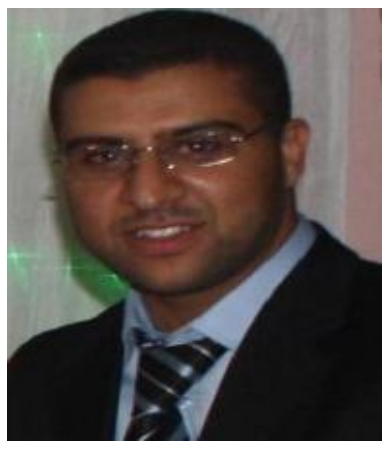

**Abdelali EL BOUCHTI** received the B.Sc. degree in Applied Mathematics from the University of Hassan 2nd, Faculty of Sciences Ain chock, Casablanca, Morocco, in 2007, and M.Sc. degree in Mathematical and Computer engineering from the Hassan 1st University, Faculty of Sciences and Techniques (FSTS), Settat, Morocco, in 2009. Currently, he is working toward his Ph.D. at FSTS. His current research interests include performance evaluation and control of telecommunication networks, stochastic control, networking games, reliability and performance assessment of computer and communication systems

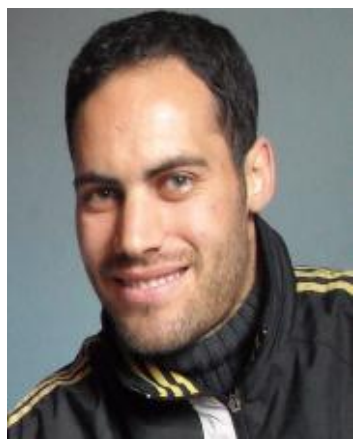

**Said EL KAFHALI** received the B.Sc. degree in Computer Sciences from the University of Sidi Mohamed Ben Abdellah, Faculty of Sciences Dhar El- Mahraz, Fez, Morocco, in 2005, and a M.Sc. degree in Mathematical and Computer engineering from the Hassan 1st University, Faculty of Sciences and Techniques (FSTS), Settat, Morocco, in 2009. He has been working as professor of Computer Sciences in high school since 2006, Settat, Morocco. Currently, he is working toward his Ph.D. at FSTS. His current research interests performance evaluation, analysis and simulation of Quality of Service in mobile networks.

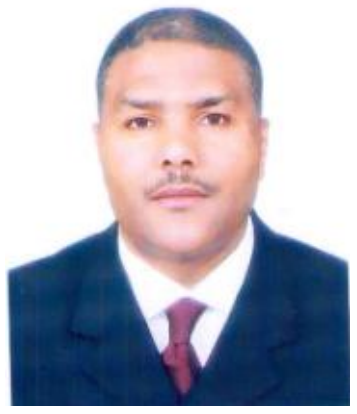

**Dr. Abdelkrim HAQIQ** has a High Study Degree (DES) and a PhD (Doctorate d'Etat) both in Applied Mathematics from the University of Mohamed V, Agdal, Faculty of Sciences, Rabat, Morocco. Since September 1995 he has been working as a Professor at the department of Mathematics and Computer at the faculty of Sciences and Techniques, Settat, Morocco. He is the director of Computer, Networks, Mobility and Modeling laboratory and a general secretary of e-NGN research group, Moroccan section. He was the chair of the second international conference on Next Generation Networks and Services, held in Marrakech, Morocco 8 – 10 July 2010.

Professor Haqiq' interests lie in the area of applied stochastic processes, stochastic control, queueing theory and their application for modeling/simulation and performance analysis of computer communication networks.

From January 98 to December 98 he had a Post-Doctoral Research appointment at the department of systems and computers engineering at Carleton University in Canada. He also has held visiting positions at the High National School of Telecommunications of Paris, the universities of Dijon and Versailles St-Quentin-en-Yvelines in France, the University of Ottawa in Canada and the FUCAM in Belgium.